# Bacteria colonies modify their shear and compressive mechanical properties in response to different growth substrates

Jakub A. Kochanowski[1], Bobby Carroll[1], Merrill E. Asp[1], Emma Kaputa[1], Alison E. Patteson[1]

1- Physics Department and BioInspired Institute, Syracuse University, Syracuse, NY

**Abstract**

Bacteria build multicellular communities termed biofilms, which are often encased in a self-secreted extracellular matrix that gives the community mechanical strength and protection against harsh chemicals. How bacteria assemble distinct multicellular structures in response to different environmental conditions remains incompletely understood. Here, we investigated the connection between bacteria colony mechanics and the colony growth substrate by measuring the oscillatory shear and compressive rheology of bacteria colonies grown on agar substrates. We found that bacteria colonies modify their own mechanical properties in response to shear and uniaxial compression with the increasing agar concentration of their growth substrate. These findings highlight that mechanical interactions between bacteria and their microenvironment are an important element in bacteria colony development, which can aid in developing strategies to disrupt or reduce biofilm growth.

**Introduction**

Living cells use physical and chemical signals from their environment to adapt their shape, size, and activity [1]. The mechanical properties of cells and tissues are often critical to their biological function, as living materials must be rigid enough to maintain their structure yet compliant enough to change shape as needed. It is now well known that many (though not all) animal cell types tune their shape and stiffness in response to changes in the mechanical properties of their local tissue microenvironment. Dysregulation of this process in humans is associated with disease [2]. Although there has been considerable study devoted to how eukaryotic cells sense and respond to physical changes in their environment, much less is known about prokaryotic systems, such as bacteria aggregates and biofilms.

Biofilms and other bacteria aggregate systems are collectives of bacteria that are typically surface bound and self-encased in an extracellular polymeric substance (EPS) matrix [3, 4]. Biofilms can have beneficial effects on ecosystems such as soil [5] and coastal environments [6], contributing to nutrient cycling and carbon balance. The resilience of multicellular bacterial biofilms, both in development and response to environmental stressors, has deleterious effects in medicine and engineering, contributing to microbial infections [7, 8] and biofouling of water ways and industrial machinery [9]. Bacteria often live in soft environments such as soils or tissues, and how they respond to physical features of those complex environments is not fully known. In general, when a bacteria comes into contact with a surface, the cell initiates a gene expression program that promotes colonization and biofilm formation [10]. The gene expression is related to EPS production via cyclic di-GMP [11], which, along with cell division and cell surface motility, drives biofilm expansion [12].

The resulting biofilm can be described as a composite biomaterial of rigid bacteria cells (colloid) in a crosslinked EPS polymer matrix (hydrogel). The EPS matrix is responsible for biofilm cohesion and architecture, including the organization of matrix-associated proteins that can mediate surface adhesion and cell-cell adhesion [13, 14]. More broadly, the EPS gives the biofilm its viscoelasticity, a property believed to be a survival response to external stresses [15]. The mechanical properties of the EPS are thought to dominate biofilm rheology, in part because the bacteria fraction in biofilms is typically small [16]. Factors such as EPS concentration [17, 18], water content [16], pH [19], and divalent cation crosslinkers [20, 21] play a role in the mechanical properties of the EPS matrix. An additional consideration is the mechanics of the substrate the biofilm grows on, which have been demonstrated to affect biofilm properties such as adhesion [22] and colonization [23]. Agar gels are a well-studied substrate for biofilm growth given their bioinert properties. Agar forms hydrogels, typically swollen with nutrient-rich media for cell studies: gels prepared with relatively low agar concentration are softer and have higher water content compared to gels prepared with higher agar concentration [12]. While it is known that bacteria growth decreases with agar concentration [24], how biofilm colony morphology and stiffness change with agar concentration is not well understood.

In this manuscript, we focus on the collective bacteria growth of *S. marcescens* and *P. aeruginosa*, a general mechanism employed by many bacteria (e.g. *E. coli*, *S. aureus*, *B. subtilis*) and fungal species (e.g. *P. chrysogenum*), on agar substrates. Here we report novel experimental data that addresses whether physical changes in bacterial growth substrate elicits physical changes in bacteria aggregates through oscillatory shear and compressive rheology. By varying agar concentration and measuring mechanical properties of collective bacteria aggregates, we find that bacteria aggregates not only change their colony size but also modify their stiffness in response to physical features of their environment. These results have important implications for understanding bacteria-materials interactions and how biofilms develop in different environments.

**Results**

### 1. *Design and characterization of bacteria colonies*

Our experimental protocol consists of culturing bacteria on agar substrates of varying agar concentration and performing rheological characterization of the resulting colonies. In this study, agar concentration is probed over a range of 1%-2%. The elastic storage modulus G′, which quantifies a material's resistance to shear deformations, of the agar gel varied from approximately 1.7 kPa to 2.5 kPa in the linear regime (See SI.) Representative images of *S. marcescens* colonies on the agar substrates are shown in Fig. 1a. The spread area of the colonies significantly decreases with increasing agar concentration from covering the entire petri dish, approximately 58 cm$^2$, on 1% agar to approximately 10 cm$^2$ on 2% agar.

To characterize the mechanical properties of the bacteria colonies, we transferred the colonies to a shear rheometer (Methods). Briefly, multiple colonies were grown up over the course of 7 days and then transferred together to form a bulk mechanical measurement. We measured their elastic storage modulus G′ and viscous loss modulus G'', which quantifies viscous energy dissipation, using oscillatory shear strain and frequency sweeps (Fig. 1). Figure 1d and f shows the oscillatory shear strain sweep of bacteria colonies over a range of strain amplitudes from 2% to 50% at a frequency of 10 Hz. We find that colonies grown on different agar substrates exhibit similar viscoelastic solid rheology behavior albeit with different magnitudes of shear modulus G' and loss modulus G''. The bacteria colonies exhibit rheological properties resembling that of a viscoelastic solid, similar to prior biofilm experiments [18, 25, 26]. For each concentration of agar the G' values of the colonies grown on them are approximately constant at small strains until a critical strain (approximately 5-10%), above which G' rapidly decreases, indicative of colony yielding. For small strains, the elastic modulus G' values are nearly 10x larger than the viscous modulus G'' (Fig. 1d,f). The G'' curves initially rise with increasing strain, then decrease above the critical strain value. SI Figure 1 shows the frequency sweeps performed at 2% strain over a range of frequencies. The data show an approximately constant G' that increases slightly with increasing frequency, indicating the bacteria colonies are behaving as viscoelastic solids at low shear strains.

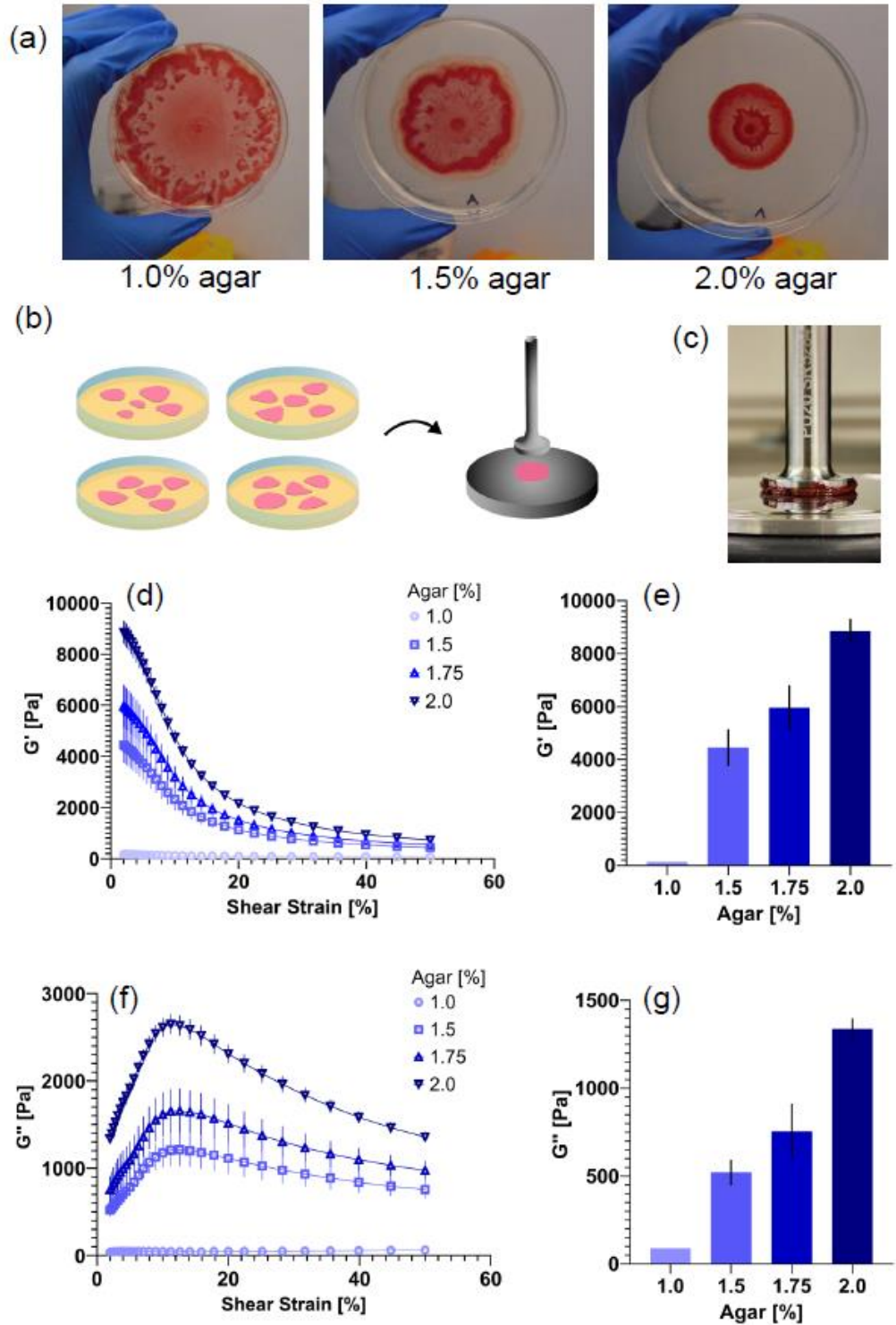

**Figure 1. Mechanical characterization of S. marcescens colonies. (a)** Representative images of *S. marcescens* colonies grown on 1.0%, 1.5%, and 2% agar. **(b)** Multiple colonies are grown on agar, then transferred to the rheometer plate for measurements, as shown schematically. **(c)** Snapshot of *S. marcescens* colony between parallel plates of the rheometer. **(d)** Average storage modulus G' as a function of shear amplitude for colonies grown on agar substrates of varying agar concentration. **(e)** The storage modulus magnitude of the colonies, defined as the average shear modulus at 2% strain, increases from approximately 130 to 9000 Pa, as the agar concentration of the growth substrate increases. **(f)** Average loss modulus G'' as a function of shear amplitude. **(g)** The loss modulus magnitude of the colonies, defined at 2% strain, increases from approximately 90 to 14000 Pa. Data are presented as a mean value ± standard error of mean (SEM).

To quantify the effects of the growth substrate on the colony mechanical properties, we next compared the low-strain shear modulus of colonies grown on agar substrates of varying concentration (Fig. 1e,g). Here, we define the low-strain shear modulus $G'_0$ and the low-strain loss modulus $G''_0$ from these data from the approximately linear regime at low strain (2%). As shown in Fig. 1e, the plateau shear modulus $G'_0$ increases from approximately 130 to 9000 Pa as agar concentration increases from 1% to 2%. The loss modulus $G''_0$ also increases from approximately 90 to 1400 Pa (Fig. 1g). Interestingly, the increase in colony stiffness is approximately 2-fold greater than the increase in stiffness of the underlying agar substrates, and the colony stiffness increases at a faster rate than the agar stiffness with increasing agar concentration (SI. Fig 2). These data indicate that for a range of agar gel there is a quantitative adaptability of the bacteria colony stiffness and rheological properties to its underlying soft growth substrate.

### 2. *Serratia colonies exhibit compression stiffening behavior when grown on stiff substrates but not soft substrates*

Next, biofilm resistance against compressive forces were measured during uniaxial compression in the parallel plate rheometer (Methods). The compression test is a sequence of 10% axial compressions, which are held for 3-minute intervals each. Throughout the test, a simultaneous oscillatory shear is applied at 2% strain and 1 Hz frequency to monitor the evolution of the biofilm's rheological response.

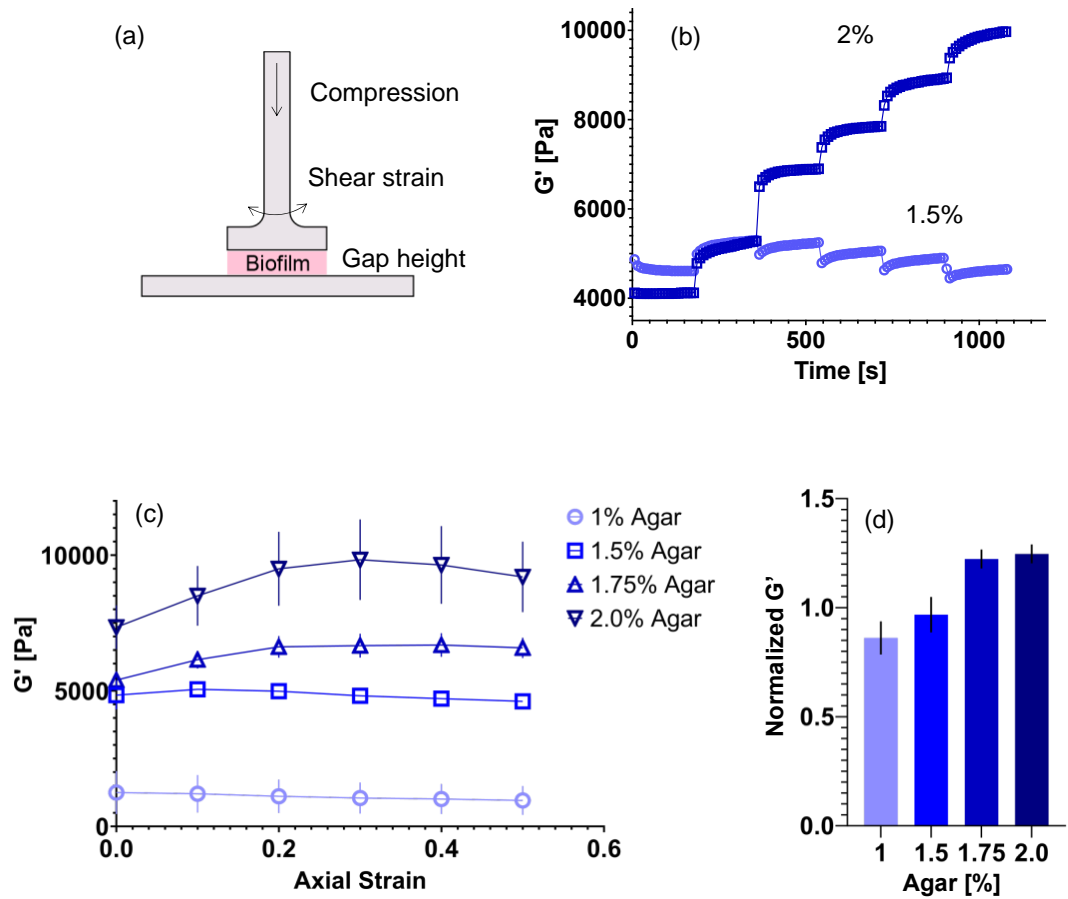

**Figure 2. Uniaxial compression of *S. marcescens* colonies. (a)** Schematic of compression test performed in a parallel plate rheometer. Uniaxial compression is applied by successively lowering the gap height between the plates. **(b)** Representative storage modulus G' over time during a compression test for S. marcescens colonies grown on 1.5% and 2.0% agar. A compressive strain of 10% is applied every three minutes. **(c)** Mean shear storage modulus G' as a function of compressive strain for *S. marcescens* colonies grown on 1.0%, 1.5%, 1.75%, and 2% agar. **(d)** The normalized compressed shear modulus ratio, defined as the final mean G' at 50% divided by the starting G' at 0% compression, increases from approximately 0.85 to 1.25, as the agar concentration of the growth substrate increases. Data are presented as a mean value ± standard error of mean (SEM).

Figure 2 shows representative compression sequence data for biofilm colonies grown on 1.5% and 2% agar substrates. While the initial uncompressed G' values of the colonies are relatively close, the response of the colonies to the stepwise compression is strikingly different (Fig. 2a). In particular, the colony grown on 2% agar shows a stepwise increase in G' with each increase in axial compressive strain: the colony's G' value increases from approximately 4000 Pa to 10,000 Pa over the 50% compression. Such rheological behavior can be interpreted as the biofilm increasing its stiffness as it is increasingly compressed, which we label here as a 'compression-stiffening' behavior. In contrast, the colony grown on 1.5% agar - while it shows a slight increase in G' after the first 10-% compression - then shows a subsequent stepwise decrease with each compressive step. This colony exhibits a 'compression-softening' behavior, at least in the regime 10-50% axial strain. For the colony grown on 1.5% agar, the final G' value at 50% compression is approximately the same as the G' value in its initial uncompressed state.

To quantify the effects of compression on the colonies, we computed the mean $G'_P$ value for multiple colonies at each compressive step. The $G'_P$ value is defined as the plateau $G'$ value for each compressive step. Fig. 2c shows the mean $G'_P$ for colonies grown on 1, 1.5, 1.75, and 2% agar. Averaging over multiple colonies, we find that there seems to be a gradual shift in the transition from compression-softening for colonies grown on low agar (1%) to the compression-stiffening behavior seen for colonies grown on 2%. The degree of compression-stiffening was quantified here by using a normalized $G'$ value (Fig. 2d), computed as the final mean $G'_P$ at 50% compression divided by the starting $G'_P$ at 0% compression for each condition. This normalized value increased from 0.85 for the 1% agar condition (compression-softening) to 1.25 for 2% (compression-stiffening).

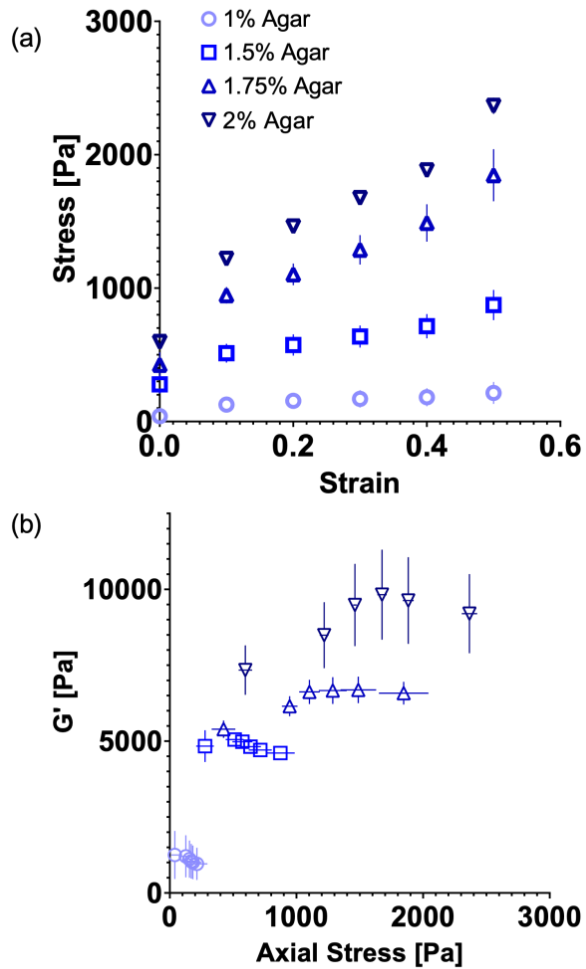

**Figure 3. Axial stress response of *S. marcescens* colonies upon uniaxial compression. (a)** The mean axial stress increases as the compressive strain increases from 0 to 50% for *S. marcescens* colonies grown on 1.0%, 1.5%, 1.75%, and 2% agar. **(b)** The average storage modulus $G'$ of *S. marcescens* colonies as a function of axial stress. Data are presented as a mean value ± standard error of mean (SEM).

To determine whether these rheological behaviors were unique to *Serratia marcescens* or whether they were shared by other bacteria species, we repeated the compression experiments with *Pseudomonas aeruginosa* (SI Fig. 3). Interestingly, we found that *P. aeruginosa* colonies exhibited similar mechanical behaviors upon compression that varied with the concentration of agar of their growth substrate. In particular, *P. aeruginosa* colonies grown on 2% agar exhibit compression-stiffening, whereas colonies grown on 1% agar exhibit a shear modulus that remains approximately constant.

Similar increases in biofilm stiffness with increasing compressive loading have recently been reported (24). It had been argued that compression drives rearrangement of cells in the colony matrix, driving contact forces between neighboring cells, and increasing the mechanical resistance of the biofilm colony. In our experiments, we find that compression-stiffening behavior depends on the growth substrate and, while compression-stiffening behavior occurs on hard 2% agar, it fades away for colonies grown on softer less-concentrated agar substrates (1.0% and 1.5% agar). Here, we will examine the effect of the growth substrate on the colony mechanics. Namely, the loss of compression-stiffening can be due to swelling of the biofilm matrix on soft agar substrates.

### 3. Axial stress response upon uniaxial compression

Next, we examined the axial stress response of bacterial colonies upon uniaxial compressive strain (Fig. 3). The axial stress of the bacterial colony is monitored in the parallel plate rheometer as the colony is subjected to increasing levels of compressive strain. Fig. 3a shows the mean axial stress (σ) versus compressive strain (ε) data for colonies grown on agar plates of varying concentration. The data are used to compute an apparent Young's modulus (E) as the slope of σ vs. ε. The apparent Young's modulus varies over an order of magnitude, rising from approximately 300 Pa on 1% agar to 3,000 Pa on 2% (Table 1). Figure 3b shows the G' value as a function of the mean uniaxial stress for colonies grown on 1, 1.5, 1.75, and 2% agar. For small values of stress, the relation between G' and uniaxial stress is approximately linear. The linear relations observed here are common feature of living materials, such as biofilms [26] and tissues [27, 28] as well as inert materials, such as rubber [29], which exhibit increases in G' upon compression (e.g. compression-stiffening behavior). Here, we find that for colonies grown on relatively more concentrated agar plates (1.75% and 2%), G' increases with uniaxial pressure with a positive slope, whereas colonies grown on less concentrated agar plates (1% and 1.5%) shift to a negative slope, though maintaining a linear G' v. σ relation. Some of the curves (agar % 1.75 and 2) hint at a transition from a linear relation with one slope to another slope value at higher uniaxial pressures. The data in Fig 3b is fit to a linear relation to obtain a slope (Table 1), focusing the fit on the initial G' v. σ linear domains at lower σ. For the colonies that stiffen upon compression, the slope is approximately 1.5 for 1.75% agar and 2.3 for 2% agar. For the other colonies, the slope is approximately -1.4 for 1% agar and 0.14 for 1.5% agar.

| Agar (%) | Apparent Young's Modulus (Pa) | G' vs Axial Stress Slope |
|---|---|---|
| 1.0 | 270.0 ± 57 | -1.4 |
| 1.5 | 1359 ± 320 | 0.14 |
| 1.75 | 3029 ± 208 | 1.5 |
| 2.0 | 2965 ± 530 | 2.3 |

**Table 1.** Apparent Young's Modulus and G' vs. Axial Stress Slope for *S. marcescens* colonies grown on growth substrates of varying agar concentration.

### 4. Substrate agar content affects biofilm volume and dry weight composition.

Our results thus far show that the mechanical properties of *S. marcescens* and *P. aeruginosa* colonies depend on concentration of agar in the growth substrate and colonies grown on stiffer, more concentrated agar compression stiffen, whereas those grown on softer, less concentrated agar do not. To interpret these results, we suggest a mechanism supported by our experimental observations that will impact the compression stiffening behavior of biofilms. While cells may have biological responses through changes in gene expression to different substrates, here, we propose an alternative physical process that could act in parallel of gene expression changes. Namely, mechanical changes in a colony's environment drives physical remodeling of colony matrices, driving changes in biofilm mechanics, in particular via the hydrogel swelling response of the colony matrix.

The impact of colony matrix swelling on colony expansion has been documented in prior studies [24, 30], which revealed a biofilm matrix takes in or lets go of water depending on an osmotic gradient between the colony and its agar substrate. The source of the osmotic pressure difference is generated by the excretion of extracellular polymers or other small molecules that act as osmolytes. Gradients in osmotic pressure draw up fluid from the agar hydrogel substrate into the biofilm, which allows the biofilm to expand. On softer less-concentrated agar substrates, the agar matrix pore size is relatively larger than in more concentrated agar substrates, allowing more fluid to flow into the colony in response to the osmotic gradient. This causes the colony to swell more on less-concentrated agar substrates compared to more-concentrated ones, which we hypothesize drives changes in the mechanical properties of the bacteria colonies here.

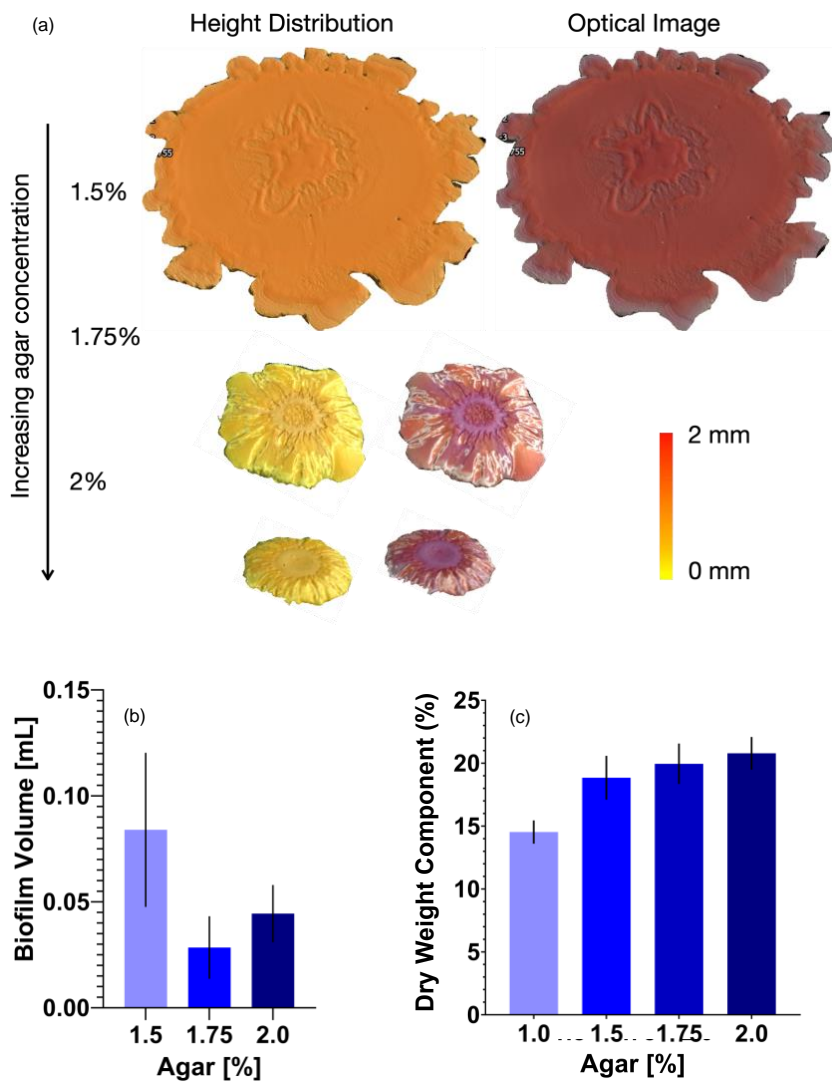

**Figure 4. Changes in *S. marcescens* colony structure. (a)** Representative optical profilometry height maps and corresponding optical images of *S. marcescens* colonies. **(b)** The volume of *S. marcescens* colonies decreases with increasing agar concentration of the growth substrate. **(c)** The dry mass content of *S. marcescens* colonies increases with increasing agar.

To connect the colony mechanics to the biofilm structural properties, we performed measurements of colony volumes and dry weight content for *S.*

*marcescens* colonies grown on varying agar substrates (Fig. 4). To quantify colony volumes, we used an optical profilometer (Keyence VR-6200) to non-invasively map out the shape of colonies on agar substrates. Figure 4a shows representative reconstructions of colonies grown on 1.5%, 1.75% and 2% agar. Here, colonies were grown for 3 days, with colonies grown on 1% agar omitted due to their tendency to overgrow 1% agar plates quickly. We found that the mean colony volume decreased from approximately 80 µL on 1.5% agar to 40 µL on 2.0% agar (Fig. 4b). Next, we estimated the dry weight content of colonies by measuring the mass of colonies before and after drying under vacuum for 24 hr at 50 °C (Methods). Figure 4c shows that fraction of dry weight content of the *S. marcescens* colonies. Here, the dry weight content of the colony is a combination of dry bacteria remains as well as the dry component of EPS matrix. We found that the percentage of dry mass increased with increasing agar concentration, rising from approximately 15% for colonies on 1% agar to 21% for colonies grown on 2% agar, consistent with the effects of increasing agar concentration as seen in recent studies of *E. coli* colonies [31]. Taken together, these data show a reduction in *S. marcescens* colony volumes on more concentrated agar substrates and an increase in dry weight fraction. These results point to increased matrix swelling on less concentrated agar substrates with higher water content, diluting the colony, and decreasing the shear stiffness of the colony.

5. **Discussion**

Previous studies have shown that altering the underlying substrate of growing biofilms can lead to large changes in how the colony spreads [24, 32-34], the forces by which colonies pull on the substrate [12, 23], and changes in gene expression profiles [35]. Here, we systemically investigated the mechanical behavior of *S. marcescens* and *P. aeruginosa* colonies grown on agar substrates of varying concentration. Our results show that for a range of agar concentrations from 1-2%, the bacteria colonies adjust their average stiffness, increasing their stiffness with increasing agar concentration. Using oscillatory and compressive rheology tests, we also found that the substrate agar concentration modified a switch between compression-stiffening to compression-softening colony behavior upon decreasing agar concentration. Finally, we have shown that the agar substrate concentration modulates colony size and dry weight fraction, which implies a change in colony structure that depends on the colony substrate.

The structural and mechanical properties of multicellular bacteria colonies are quite complicated. By measuring colony stiffness, volume, and dry mass, we quantified the morphological and structural properties of *S. marcescens* colonies as a function of the concentration of agar they were grown on. Using an optical profilometer to map colony shapes, we found that the volume of colonies decreased with increasing agar. We also found that the dry weight fraction of colonies increased with agar concentration. These findings are consistent with a substrate-dependent mechanical model of biofilms grown on agar defined by the hydrogel swelling properties of agar and the colony matrix [30, 31]. Recent work on biofilm colonies pointed out the importance of osmotic swelling in colony spreading and growth. These studies have shown biofilm-producing cells release extracellular proteins that act as osmolytes, generating an osmotic gradient between the bacteria colony and its agar substrate. This leads to a net fluid flow from the agar substrate into the bacteria colony, allowing the colony to take up fluid mass by swelling. The response of the colony to a more concentrated agar substrate involves taking up fluid through a denser agar substrate, which hinders flow and decreases the ability of the colony to swell for the same osmotic gradients [24, 30]. From this perspective, colonies on less concentrated agar swell more, take up more volume, and have a lower effective EPS polymer and cell density. Consistent with this view, our results show a decreased colony volume and higher dry weight fraction for colonies grown on more concentrated agar substrates. One outcome of these changes is that a higher concentration of extracellular polymers and cells increases a colony's resistance to shear deformation, consistent with the stiffer colonies on more concentrated agar substrates (Fig. 1).

Here, we show that the compression-stiffening behavior of biofilms is modified by the growth substrate and that compression-stiffening does not occur for biofilms grown on soft agar substrates. An emerging number of investigations is directed at understanding the compressive-stiffening behavior of biological materials [27, 28, 36-40]. Thus far, most work has focused on mammalian cells and tissues. Interest has been spurred on by the observations that many tissues, including fat, liver, and brain, stiffen upon compression, however networks comprised of the biological polymers that comprise them *soften* under compression. Finding appropriate models to capture the mechanical transition between a biological fiber network and whole cells and tissues has been an interesting material science and engineering problem. Experimentally, a compressive-softening biological fiber network can be converted to a compression-stiffening network by the addition of volume-conserving cells or particles embedded into the network [28, 36]. Several computational models have been developed to capture this mechanical response, revealing different physical mechanisms for compression-stiffening behavior. These mechanisms include: (1) deformations of the network induced by deformations of soft particles in the network, (2) heterogenous strain of the network arising from relative displacements of the particles, (3) area and volume constraints in the network that induce network bending, and (4) compression-induced jamming of the particles inside the network.

Recent studies on biofilms have considered their stiffening response to uniaxial compressive loading, so-called compression-stiffening behavior. Lysik *et al* reported compression stiffening in biofilms created by *Pseudomonas aeruginosa*, *Staphylococcus aureus*, and *Candida albicans* grown on glass surfaces in a nutrient-rich bath [40]. It was argued that increasing cell density give rise to biofilm compression stiffening, by increasing the contact between cells as a colony is compressed. In our experiments, colonies were grown at an agar-air interface. The typical volume fraction of gram-negative bacteria in colonies grown on agar is 10% or less [41, 42], below volume fractions for significant levels of compression stiffening predicted for biopolymer-cell networks [37, 38].

Here we propose a minimal model to explain our data showing a switch from biofilm stiffening to softening with decreasing agar concentration of the growth substrate based on two physical ingredients. Namely, (1) significant matrix crosslinking providing angular-constraining crosslinking and volume constraints that give rise to compression-stiffening in a network (model #3), and (2) osmotic swelling of biofilms that dilute the EPS network and crosslinking components (Fig. 5). The EPS is largely comprised of long polysaccharide polymers, such as alginate, cross-linked together by specific cell-released matrix-associated proteins and non-specific divalent cation interactions. Biofilm polysaccharides are also highly charged negative polyelectrolytes that interact strongly with divalent cations including calcium and magnesium, which serve as gelling agents to form strong hydrogels from the negatively charged polymers released by cells forming a biofilm [43]. A polyelectrolyte network of DNA becomes stiffer with higher concentrations of divalent cations [26]. Extracellular DNA is a major component of the EPS networks, providing a structural scaffold for the colony and enhancing biofilm adhesiveness [44]. Divalent cations also have significant effects on the compressive behavior of networks. Lysik et al also found that addition of highly concentrated cations could switch DNA solutions to a compression-stiffening regime. A switch to a compression-softening regime could arise from a lower concentration of angular-conserving crosslinks that create volume conserving polymer loops [37] that could result from significant swelling of the biofilm matrix on low-agar concentration substrates. Interestingly, swelling of the matrix would also lower EPS fiber density, a parameter that has yet to be systemically studied in compression-stiffening fiber network models. Taken together, these data suggest that compression stiffening in biofilms may arise from the effects of additional crosslinks in the EPS matrix, which could occur from cations released by cells in addition to the inclusion of volume-conserving cells and adhesive contact between microbial cells and the EPS network.

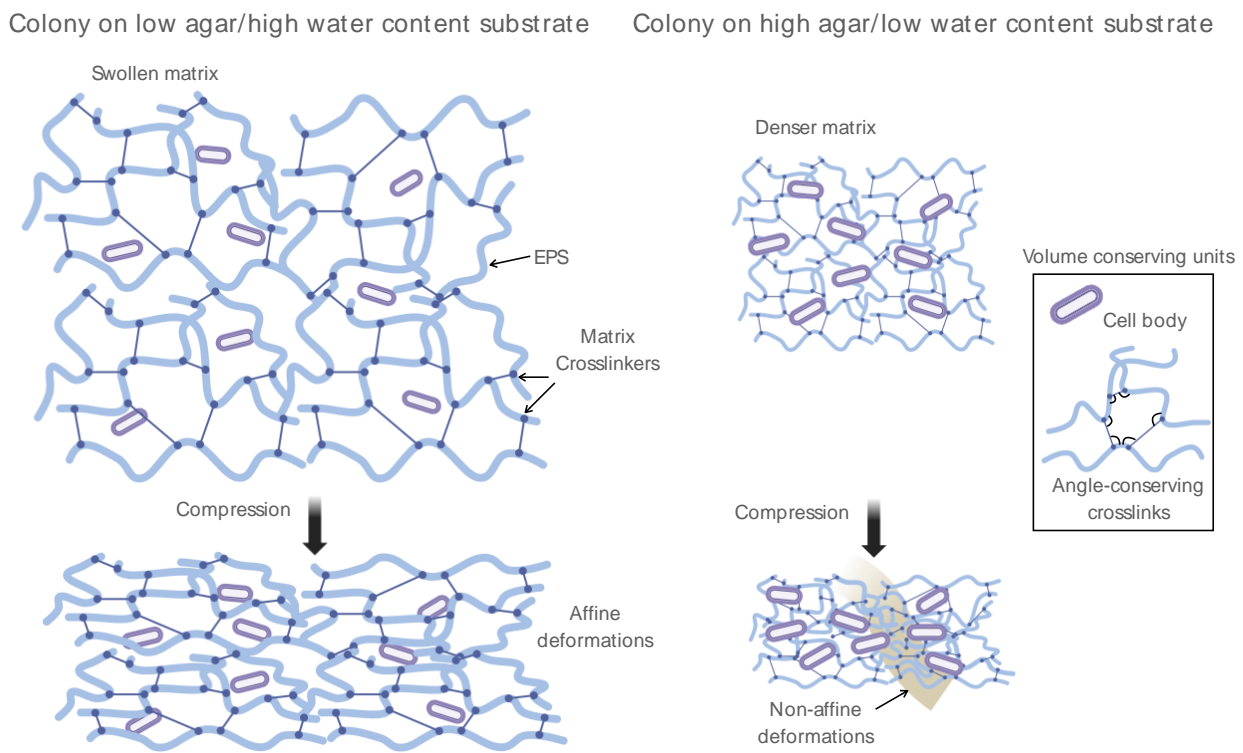

**Figure 5. Schematic model of colony mechanical transition.** Biofilm colonies grown on high water content hydrogel substrates (low agar concentration) are larger and have a higher water content themselves compared to colonies grown on low water content substrates (high agar concentration). Colonies grown on low water content substrates are denser with a higher dry-weight percentage coming from cells, EPS polymers, and other matrix-associated proteins and molecules that crosslink the network together. Upon uniaxial compressive loading, a higher density of cells and other volume conserving units (such as polymer loops generated from a higher density of crosslinks) resist deformations, driving non-affine deformations and remodeling in regions throughout the network, leading to a compression-stiffening effect.

### 6. Conclusion

To conclude, we used *Serratia marcescens* as a model bacterium to investigate the connection between bacteria colony mechanics and the colony growth substrate. The results presented here assert that the physical properties of a bacteria colony's growth substrate are a critical regulator of the colony stiffness. Bacteria colonies increase their own stiffness with increasing stiffness of their agar growth substrate. Further, bacteria colonies can switch between compression-stiffening and compression-softening behavior, depending on the concentration of their agar substrate, likely due to changes in water content of the bacteria colonies. The understanding gained here highlights that mechanical interactions between bacteria and their microenvironment are an important element in bacteria colony development. These interactions and their emergent feedback mechanisms are crucial to many issues in engineering, biology, and medicine, such as means to enhance or disrupt biofilms on different surfaces.

**Methods**

***Bacteria culture.***

Bacteria cultures of *S. marcescens* (274 ATCC) and *P. aeruginosa* (Xen05) are prepared as follows. Bacterial cells were inoculated and grown in LB medium at 37°C overnight at a shaking speed of 200rpm. For all measurements, 5 µl of inoculum was spotted on agar growth substrates. Cell plates were then maintained at 37°C for up to 7 days. *Pseudomonas aeruginosa* Xen05 was kindly provided by Dr. Robert Bucki (Medical University of Bialystok).

***Biofilm extraction via manual scraping.***

Biofilm samples were prepared with 4-5 inoculation points on each petri dish and allowed to grow for 7 days. Each measurement sample consisted of 4-5 petri dishes worth of biofilms. To transfer to the rheometer plate, samples were extracted via manual scraping. Scraping was done with the flat edge of a polyurethane rubber sheet, gently scraping along the agar surface to extract the biofilm colonies. The collective total biofilm mass after scraping is transferred to the rheometer plate for sample measurement.

***Rheological characterization***

All rheology measurements were performed on a Malvern Panalytical Kinexus Ultra+ (Malvern Panalytical) rheometer using a 20 mm parallel plate geometry at 25°C. The gap height varied based on sample amount but was maintained at approximately 1 mm. Frequency sweeps are performed at 2% shear strain amplitude at a frequency range of 0.063 rad/s to 314.2 rad/s. For shear amplitude sweep tests, the shear modulus was measured as a function of shear strain from 2% to 50% at a frequency of 1 Hz. All compression tests were performed by applying a continuous oscillatory torque at 6.3 rad/s and 2% shear strain. During compression tests, samples were subjected to stepwise compressive strains, between which samples were measured continuously for 3 min. The gap height was lowered in steps of 10% compressive strain, up to 50% strain.

***Biofilm dry weight quantitation.***

Bacteria culture was prepared to produce 14 plates of each agar % with 3 equidistant inoculation points per petri dish. The biofilms were extracted via manual scraping and combined into small petri dishes for each agar %. The biofilms in petri dishes were dried at high vacuum at 50°C for 24 hours to fully dehydrate the biofilm sample and leave only the solid fraction. The solid fraction is calculated by the following equation:

$$S(\%) = \frac{w_{dry} - w_{pd}}{w_{hydrated} - w_{pd}},$$

with $w_{dry}$ being the dry weight of the sample, $w_{pd}$ being the weight of the small petri dish, and $w_{hydrated}$ being the hydrated weight of the sample. All weight measurements were performed several times using a high precision scale.

**Acknowledgements**

We thank J.M. Schwarz, Paul Janmey, and Robert Bucki for insightful discussions. We acknowledge Austin Gardner and Sierra Weil for help assisting in profilometry experiments and photographs from Miranda Azemi. A.P. acknowledges funding from NSF MCB 2026747, NSF DEB 2033942, and the Research Corporation for Science Advancement's award CS-CSA-2023-097.